# ELEVATED SOIL LEAD: STATISTICAL MODELING AND APPORTIONMENT OF CONTRIBUTIONS FROM LEAD-BASED PAINT AND LEADED GASOLINE

By R. Dennis Cook and Liqiang Ni

*University of Minnesota and University of Central Florida*

While it is widely accepted that lead-based paint and leaded gasoline are primary sources of elevated concentrations of lead in residential soils, conclusions regarding their relative contributions are mixed and generally study specific. We develop a novel nonlinear regression for soil lead concentrations over time. It is argued that this methodology provides useful insights into the partitioning of the average soil lead concentration by source and time over large residential areas. The methodology is used to investigate soil lead concentrations from the 1987 Minnesota Lead Study and the 1990 National Lead Survey. Potential litigation issues are discussed briefly.

**1. Introduction.** Lead poisoning is a major concern, particularly for children living in older urban environments. In addition to air, household dust, food and water, soil is part of a complex system of pathways along which children can be exposed to lead [49]. Lead-contaminated soil, one of the lead reservoirs that contributes to urban environments, has been the focus of numerous studies, including investigations into the potential efficacy of soil lead abatement [1, 9, 18, 36]. The Environmental Protection Agency recently set soil lead guidelines, along with guidelines for dust and paint [51].

It is crucial to understand the contributions of pollutant sources in order to reduce risk, to mitigate impact and to allocate responsibility. Most source apportionment methods are developed from receptor-based models [24, 25]. Recently, Park et al. [40] studied multivariate receptor modeling using MCMC. Graney et al. [22] investigated the relative importance of local sources of mercury in aerosols from urban areas in south Florida using a multi-element tracer approach. Christensen and Gunst [8] studied measurement error models in chemical mass balance analysis of air quality data.

In this article we develop a novel nonlinear regression model, which is essentially a chemical mass balance model, for soil lead concentrations over time.

Despite numerous sources of soil lead [14], it is generally recognized that lead-based paint, lead aerosols from automotive emissions and point source emitters such as mineral processing plants and secondary lead smelters [2, 28] are primary sources of elevated lead concentrations in urban soils [4, 50]. Lead-based paint and automotive emissions are often cited as primary causes of elevated soil lead around urban residences, although conclusions regarding their relative contributions are mixed [50].

*Paint.* Lead-based paint was widely used on both interior and exterior surfaces until it was virtually banned by the late 1970's [47]. Browne and Laughnan [3] estimated that, under conditions of normal weathering, lead-based paint on exterior surfaces of the type used in the 1930's and 1940's eroded at an average rate of about 0.5 mil/year, contributing lead to nearby soil. Based on a study of painted rural farmhouses, Ter Haar and Aronow [45] concluded that paint lead may contribute to elevated soil lead levels 10 feet or more from the house. Davis and Burns [15] found that storm water runoff from exterior surfaces with lead-based paint is a significant source of soil lead, while lead contributions from roof rain, a conceivable transporter of aerosol lead, are relatively minor.

Soil around the foundation of a residence is widely reported to have higher lead levels than soil at remote locations [42, 48]. Foundation and yard soils around structures with brick exteriors have, on the average, significantly lower concentrations than those from houses with painted siding [42]. The 1998 EPA literature review [50] on sources of lead in soil included eight studies that reported significant associations between soil lead and exterior paint lead variables. It was concluded in the review that "...higher paint-lead loadings on exterior surfaces are associated with increased lead concentration in the surrounding soil." Francek [21] studied soil lead in a small urban area and as a result suggested that lead-based paint on older houses contributes lead to the surrounding soil.

*Gasoline.* Lead was added to gasoline from the mid 1920's until the mid 1980's [29]. Starting in the early 1970's, it was phased out because of documented adverse health effects. Many studies have documented the increase in soil lead adjacent to busy roads [23]. Singer and Hanson [44] found that the elevated soil lead levels (128 to 700 ppm) adjacent to highways in the Minneapolis–St. Paul metropolitan area were related to traffic volume and distance from the highway.

Other studies focused on generalized lead deposition over urban areas [26, 42]. Significant associations have been found between ambient air lead levels and soil lead in studies conducted while leaded gasoline was still prevalent.

Page and Ganje [38] estimated that, over a period of 40 years in the Los Angeles metropolitan area, between 15 and 36 ppm of lead accumulated in the surface 2.5 cm of soil. Tiller et al. [46] studied the regional distribution of lead from surface soils around Adelaide, Australia. They concluded that aerosol lead from the metropolitan area can spread up to 50 km from the center of the city.

Mielke and others [32, 33, 34] have argued in favor of the *aerosol hypothesis*: lead aerosols from automotive emissions collect on exterior surfaces and are subsequently washed into surrounding soil by precipitation. Most paint lead likely remains attached to the surfaces where it was applied, and leaded gasoline is substantially responsible for the elevated lead levels observed in soils near building foundations. This aerosol hypothesis seems consistent with the finding by Chaney and Mielke [6] that high concentrations of lead in garden soil were found in areas of predominantly unpainted brick buildings.

*Concomitant variables.* Evidence for the culpability of lead-based paint and leaded gasoline as contributors to soil lead is based in part on using statistical methods, particularly linear regression and analysis of variance, to characterize the relationship between soil lead and local and area variables such as distance from structure, lead-based paint loadings determined using X-ray fluorescence, siding type, condition of paint, presence of paint chips, distance to nearest road, community traffic patterns and structure age. Of such concomitant variables, structure age is often cited as perhaps the strongest single predictor of lead concentrations in nearby soil [21, 48, 50].

Nevertheless, structure age is only a surrogate for the accumulation of soil lead over time. Soil lead is quite immobile and tends to stay near the surface [52]. Analysis of historical lead smelting sites [31] showed vertical migration is not significant. Most soil lead remains in the top few inches after a long time, even a few centuries. Thus, it is reasonable to conclude that lead from paint and gasoline accumulated in residential soils.

*Objectives.* The objectives of this article are two fold: First, a new nonlinear regression model is proposed for statistical analysis of soil lead. The model is based on using cumulative lead exposures (defined herein) as the main predictors instead of structure age. This model has the advantage of tying soil lead concentrations directly to sources of paint and gasoline lead, and has the potential to permit apportionment over large residential populations. Second, the model is used to analyze data from the 1987 study conducted by the Minnesota Pollution Control Agency and data from the 1990 National Survey sponsored by the US Department of Housing and Urban Development. While the proposed model can be used with local variables, only national and regional cumulative exposure variables are used in this report to explain the variation in soil lead concentrations over time and to partition the concentrations by source.

**2. Model development.** The proposed model is based on a conceptual framework for characterizing the concentration $L_y$ of lead in a soil sample taken in year $Y$ near a structure built in year $y \leq Y$. We suppress $Y$ in notation, assuming for simplicity that all samples were taken in the same year. Soil lead contamination is additive [4, 50] so that, as a conceptual starting point, $L_y$ can be represented as

$$L_y = \beta + \rho_y + \gamma_y. \tag{2.1}$$

In this equation $\beta$ represents a random "background" concentration that would still be present in the absence of contributions from paint and gasoline. Its average magnitude will depend on the study site since natural soil lead levels vary by location, with a national geometric mean of about 16 ppm [50, 52]. It can be inflated by other sources, like lead from nearby solid waste incineration [7], for example. The term $\rho_y$ represents the part of the concentration that is attributable to lead-based paint and $\gamma_y$ represents the part attributable to leaded gasoline.

The immobility of lead in soil suggests that cumulative lead exposure is more relevant than current exposure to understanding soil lead concentration. Thus, on the average, we expect that the magnitude of the terms $\rho_y$ and $\gamma_y$ will increase with age and be relatively large for older residences. This is consistent with the common finding that, on the average, total concentration $L_y$ in the soil around residences increases with their age.

The paint component $\rho_y$ depends on the amount of paint applied to a structure between the year $y$ in which it was built and the study year $Y$. Let $\{y_1, y_2, \ldots, y_m\}$ denote the years in which a structure was painted and let $p_i$ denote the amount of lead in the paint applied in year $i$. Then $\rho_y$ can be expressed as

$$\rho_y = \sum_{i=1}^{m} f_{y_i} p_{y_i},$$

where $f_{y_i}$ represents the fractional loss of lead to nearby soil from paint applied in year $y_i$. This additive representation is reasonable since, after new paint is applied, the amount of lead contributed to the soil from previous painting should be quite small. The $f_{y_i}$'s can depend on a variety of factors, including the side of the structure on which the sample was taken (in the northern hemisphere, southern exposures tend to deteriorate most rapidly), whether old paint was removed prior to painting, chalking, weathering and quality of the paint. In the absence of conditioning information to identify such factors, it is reasonable to average the fractional losses so they are no longer structure or year specific. Assuming that the $f_{y_i}$ are independent and identically distributed, let $\bar{f} = \mathrm{E}(f_{y_i})$. Then, assuming that the $f_{y_i}$'s are independent of the amounts $p_{y_i}$, we have

$$\mathrm{E}(\rho_y | p_i, i = y, y+1, \ldots) = \bar{f} \sum_{i=1}^{m} p_{y_i} \approx \frac{\bar{f}}{c} \sum_{i=y}^{Y} p_i = F P_y,$$

where $c$ represents the average number of years between paintings, $F$ is the average yearly loss, and $P_y = \sum_y^Y p_i$ represents the cumulative exposure to lead from lead-based paint. The $p_i$'s depend strongly on the year, reflecting the changing concentration of lead in paint. Yearly amounts $p_i$ for residences built after 1980 are expected to be much smaller than for residences built in the 1920's, for example. In short, the expected contribution of lead-based paint to soil contamination around a structure built in year $y$ can be expressed to a useful approximation as

(2.2) $$\mathrm{E}(\rho_y | P_y) = F P_y.$$

This representation depends on the modeling assumption that the $f_{y_i}$'s are independent of the $p_{y_i}$'s. One can imagine scenarios leading to dependence, but we have no firm information to guide us. An extension that treats the $f$'s as a random effect is described in Section 5.

Reasoning similarly, the part $\gamma_y$ of $L_y$ attributable to leaded gasoline can be modeled as

(2.3) $$\mathrm{E}(\gamma_y | G_y) = H \sum_{i=y}^{Y} g_i = H G_y,$$

where $G_y = \sum_y^Y g_i$ and $g_i$ is the contribution of lead from leaded gasoline in year $i$. Like the yearly contributions from lead-based paint, the contributions $g_i$ from leaded gasoline depend strongly on year. For instance, $g_i = 0$ for $i \leq 1920$. We refer to $G_y$ and $P_y$ as *cumulative exposure predictors* because their role is to provide statistical information on the average exposure to lead from paint and gasoline. Operational versions of these predictors are discussed in Section 3.

Since $\gamma_y$ and $\rho_y$ are unobservable, it is not possible to estimate directly the mean functions in equations (2.2) and (2.3). However, progress is possible through the mean function

(2.4) $$\mathrm{E}(L_y | G_y, P_y) = \mathrm{E}(\beta) + \mathrm{E}(\rho_y | G_y, P_y) + \mathrm{E}(\gamma_y | G_y, P_y).$$

For an uncomplicated attribution to source, $(G_y, P_y)$ should satisfy the following two relations:

(2.5) $$\mathrm{E}(\rho_y | G_y, P_y) = \mathrm{E}(\rho_y | P_y),$$

(2.6) $$\mathrm{E}(\gamma_y | G_y, P_y) = \mathrm{E}(\gamma_y | G_y).$$

They seem reasonable since a structure's cumulative exposure to gasoline lead would seem to furnish little if any information on $\rho_y$ beyond that provided by its cumulative exposure to paint lead and vice versa.

Many studies have found that soil lead concentrations from replicate samples are skewed and likely log-normally distributed [48]. Our experiences support this conclusion. Accordingly, we incorporate a stochastic component into our model for the mean function by using the logarithm of the concentration:

$$(2.7) \qquad \log(L_y) = \log\{\mathrm{E}(\beta) + \mathrm{E}(\rho_y|P_y) + \mathrm{E}(\gamma_y|G_y)\} + \varepsilon,$$

where $\varepsilon$ is a normal random variable with mean 0 and variance $\sigma^2$. This model is an instance of transform-both-sides methodology [5], except here the transformation was taken from past studies and not estimated based on this study. Used in combination with the mean models (2.2) and (2.3), equation (2.7) is the nonlinear model we use for studying soil lead concentrations. Its formulation depends on cumulative exposure predictors and thus differs from past investigations that relied mostly on predictors measured at the time of the study.

A comparison of the relative magnitudes of the terms in equation (2.7) will likely be of interest in many studies and is of particular interest in the Minnesota and National Surveys discussed in Section 3. Comparisons can be made by estimating the *fractional contributions* $F_{(\cdot)}(G_y, P_y)$ of background, paint and gasoline defined respectively as functions of $(G_y, P_y)$ by the following equations:

$$F_\beta(G_y, P_y) = \mathrm{E}(\beta)/\mathrm{E}(L_y|G_y, P_y),$$
$$F_\rho(G_y, P_y) = \mathrm{E}(\rho_y|P_y)/\mathrm{E}(L_y|G_y, P_y),$$
$$F_\gamma(G_y, P_y) = \mathrm{E}(\gamma_y|G_y)/\mathrm{E}(L_y|G_y, P_y).$$

The lead content of paint and gasoline varied substantially from the late 1800's to the present and, thus, it should be possible by using appropriate proxies for $P_y$ and $G_y$ to develop an instance of model (2.7) that distinguishes between paint and gasoline. In this report we use national and regional lead consumptions to construct $P_y$ and $G_y$, as described following introduction of the Minnesota Lead Study and the US National Survey.

### 3. Minnesota and National Survey Studies.

*Minnesota study.* In 1985 the Minnesota Legislature directed the Minnesota Pollution Control Agency (MPCA) to study a variety of lead related issues, including the extent of lead contamination in the soil. In response, the MPCA developed a sampling plan that gave preference to census tracts consisting largely of old inner-city neighborhoods with poorly maintained housing, the kind of situation in which they expected to find relatively high levels of lead contamination [42].

We studied lead concentrations $L_y$ (µg/g) obtained from the MPCA public records for *foundation* and *yard* samples from the Twin Cities. Foundation samples were collected within 1.5 m of a residence, while yard samples were collected farther from the structure, both from the top 2 cm of soil. Structures with incomplete records or notes indicating a point source nearby were excluded. For the reasons indicated shortly, also excluded were all structures built prior to 1902. This left 132 foundation samples and 219 yard samples. The year in which a structure was built is not available in the study records. Instead, using the addresses available in the study records, these dates were determined from the public record.

*US National Survey.* Our second analysis uses data from the 1990 National Survey of Lead-Based Paint in Housing initiated by the Department of Housing and Urban Development [48]. The National Survey measured lead in the exterior soil of 381 housing units in 30 counties across the 48 contiguous states. The sampled population was designed to be representative of US housing constructed prior to 1980, and consisted of urban, suburban and rural houses. For each house sampled, soil from the top 2–3 cm was evaluated for lead at three locations: outside the main entrance, along the drip line of a randomly selected exterior wall approximately 12 inches from the structure and at a remote location approximately half way between the drip line sample and the property line.

We studied the lead concentration $L_y$ (µg/g) for the 275 structures with drip line samples and the 276 structures with remote samples. To maintain consistent terminology with Minnesota, we refer to these as the *foundation* and *yard* samples, although the sampling protocol is not quite the same in the two studies. The year built is available in the public record of the study.

Figure 1 shows scatterplots of $L_y$ in log scale versus year built $y$ for the four situations under consideration, foundation and yard samples from the US and Minnesota studies; the lines on the plots are discussed later. The distribution of points along the abscissae reflects the different objectives of the studies. The Minnesota study is dominated by inner-city housing built before 1930, while the US study is a mix of metropolitan and rural houses built mostly after 1930.

*Cumulative exposure predictors.* In both studies, we use the cumulative amount $T_y = \sum_{i=y}^{Y} w_i$ of lead in white lead pigment as a proportional proxy for the cumulative paint predictor $P_y \propto T_y$:

$$(3.1) \qquad \mathrm{E}(\rho_y|T_y) = \theta_1 T_y = \theta_1 \sum_{i=y}^{Y} w_i,$$

where $w_i$ is millions of metric tons of lead used in the production of white lead pigment in the US in year $i$ as determined from the annual volumes for 1902 to 1979 of the US Bureau of Mines *Minerals Yearbook*. Nonpaint applications of lead were excluded in the $w_i$'s. Because lead-based paint was virtually banned by 1980, we set $w_i = 0$ for $i \geq 1980$. The measurement year is $Y = 1986$ for the Minnesota study and $Y = 1990$ for the National Survey. White lead production is not available in the *Minerals Yearbook* for years prior to 1902. Lacking firm information on $w_i$ prior to 1902, we excluded all samples from structures built before that date. Thus, in model (2.7) $y \geq 1902$ for all samples. The parameter $\theta_1$ in equation (3.1), which is proportional to $F$, will be estimated by the method of maximum likelihood.

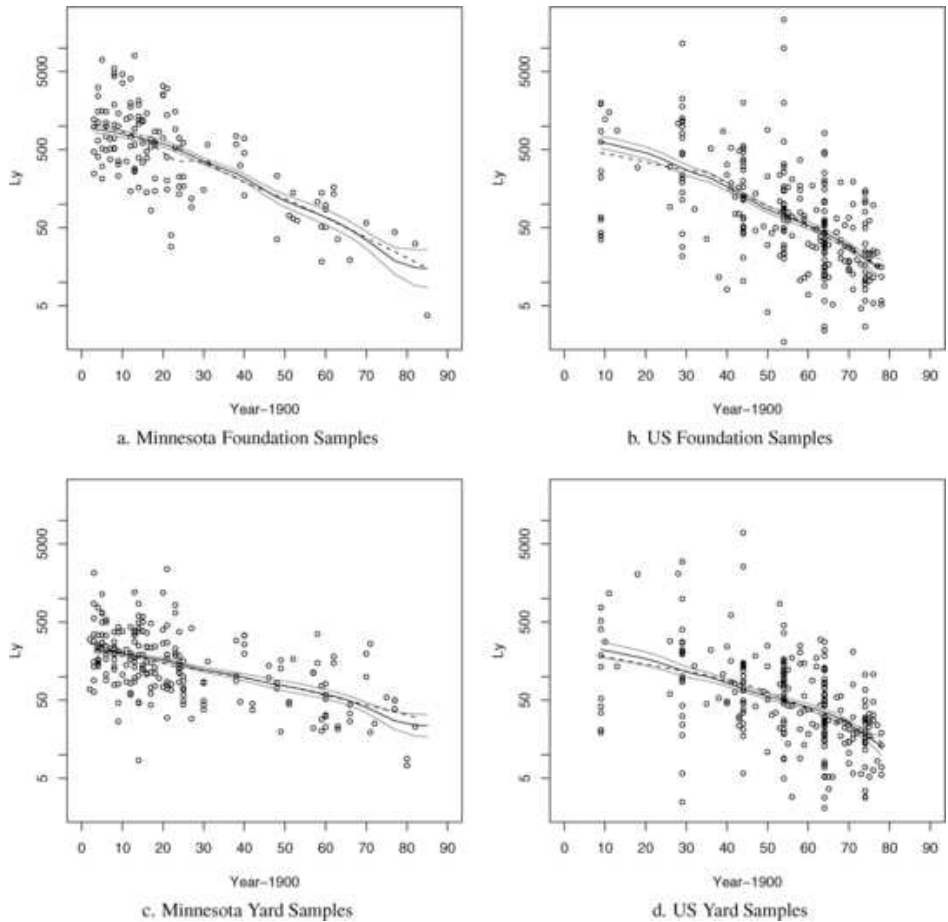

FIG. 1. *Soil lead concentration (ppm) in* $\log_e$ *scale versus year built from foundation and yard samples for Minnesota and US data. Black lines: fitted values plus and minus one bootstrap standard error. Dotted line: lowess mean with tuning parameter* 0.7.

An alternative is to use total lead in lead pigment rather than white lead. However, white lead pigment was widely used in paints for residential structures, and white lead is highly correlated with total lead over time.

Similarly, for the National Survey, we use the cumulative consumption $U_y = \sum_y^Y u_i$ of lead in gasoline as a proportional proxy for the cumulative predictor $G_y \propto U_y$:

$$\text{(3.2)} \qquad \text{E}(\gamma_y | U_y) = \theta_{U2} U_y = \theta_{U2} \sum_{i=y}^{Y} u_i,$$

where $u_i$ is millions of metric tons of lead consumed in gasoline in the USA in year $i$ [27]. Aviation fuel and gasoline sales to the military were excluded. The measurement year is again $Y = 1990$. Sales of leaded gasoline were negligible prior to 1924 [37] and, thus, we set $u_i = 0$ for $i \leq 1923$. Like $\theta_1$, $\theta_{U2}$ is an unknown rate parameter to be estimated.

For the Minnesota study, we use the scaled cumulative consumption $M_y = \sum_{i=y}^{Y} 50 m_i$ of lead in gasoline as $G_y$:

$$\text{(3.3)} \qquad \text{E}(\gamma_y | M_y) = \theta_{M2} M_y = \theta_{M2} \sum_{i=y}^{Y} 50 m_i,$$

where $m_i$ is millions of metric tons of lead consumed in Minnesota gasoline in year $i$, as determined from Ethyl Corporation's *Yearly Report of Gasoline Sales by State* [17], $Y = 1986$ and as for the US data $m_i = 0$ for $i \leq 1923$. However, direct data on $m_i$ is unavailable from 1924 to 1934. To compensate, we imputed these values of $m_i$ by using the predicted values $\hat{m}_i = 0.0205 u_i$ from the simple regression through the origin of $m_i$ on $u_i$ for $1935 \leq i \leq 1974$ ($R^2 = 0.996$); see Figure 2. Only 1935 through 1974 data was used for the imputation, because after 1974 the lead content of gasoline dropped faster in Minnesota than in US. This imputation turned out to be unimportant since results based on it are nearly identical to the results based on using $m_i = 0$ for $i \leq 1934$. Finally, the factor 50 is included in equation (3.3) to facilitate interpretation since it places Minnesota consumption on a National scale.

Using these approximations for cumulative lead exposure, we reach our operational models for the Minnesota data

$$\text{(3.4)} \qquad \log(L_y) = \log(\theta_{M0} + \theta_{M1} T_y + \theta_{M2} M_y) + \varepsilon$$

and the National Survey data

$$\text{(3.5)} \qquad \log(L_y) = \log(\theta_{U0} + \theta_{U1} T_y + \theta_{U2} U_y) + \varepsilon,$$

where $\varepsilon$ is assumed to be a normal random variable with mean 0 and variance $\sigma_M^2$ for the Minnesota data and $\sigma_U^2$ for the US data. We used these models for separate analyses of the foundation and yard samples.

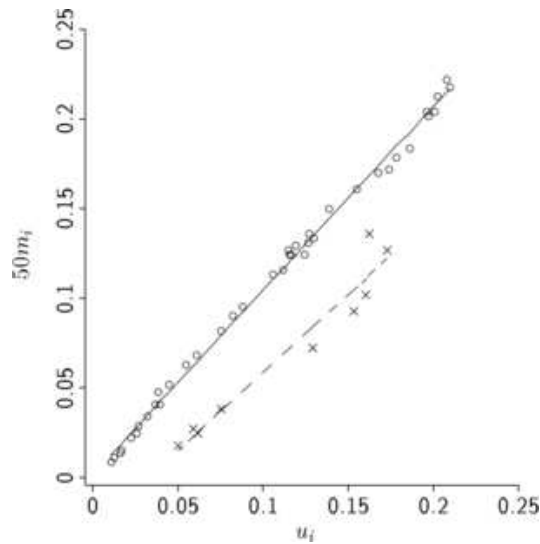

Fig. 2. *Yearly Minnesota consumption $50m_i$ of lead in gasoline versus US consumption $u_i$ with 1935 to 1974 indicated by ○ and 1975 to 1983 by ×. The best fit lines to the two groups are shown.*

There is a value for each of the cumulative predictors $M_y$, $U_y$ and $T_y$ for all years between 1902 and 1990, but only the values corresponding to years in the data are required by models (3.4) and (3.5). Figure 3 shows plots of these predictors for the years present in the Minnesota and US data; the $M_y$ points were joined for visual clarity. The shape of the Minnesota curve for cumulative exposure to gasoline lead is quite similar to that for the US, indicating that these predictors might be exchangeable. However, the predictor $T_y$ for paint lead is quite different than those for gasoline, suggesting that it might be possible to distinguish statistically between the contributions of paint and gasoline to lead in residential soils. Figure 3 supplants the need to consider the usual charts of yearly gasoline and lead consumption and reinforces the conclusion that cumulative consumption is relevant.

*Estimation and source apportionment.* Estimates $\hat{\theta}_{\cdot k}$ of the rate parameters $\theta_{\cdot k}$, $k = 0, 1, 2$, in models (3.4) and (3.5) were determined by the method of maximum likelihood. All $\theta$ parameters and their estimates are in parts per million. Thus, for example, $\hat{\theta}_{M0}$ is the estimate in ppm of the average background concentration $\theta_{M0}$ for the Minnesota data, and $\hat{\theta}_{U2}$ is the estimated increase in ppm per one million metric ton increase in cumulative exposure to lead from leaded gasoline. Similarly, for the Minnesota and US data,

$$(3.6) \qquad \hat{\mathrm{E}}(L_y | T_y, M_y) = \{\hat{\theta}_{M0} + \hat{\theta}_{M1} T_y + \hat{\theta}_{M2} M_y\} \exp(\hat{\sigma}_M^2 / 2)$$

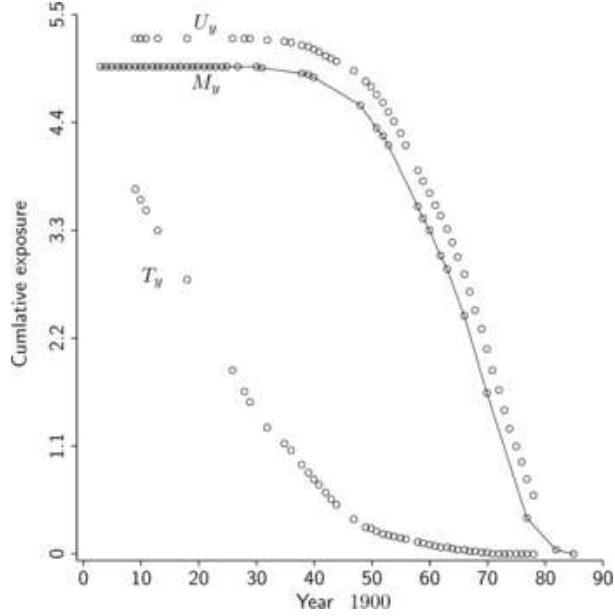

Fig. 3. *Cumulative exposures to paint lead $T_y$ and gasoline lead, $U_y$ and $M_y$, in millions of metric tons, by year.*

and

$$(3.7) \qquad \hat{\mathrm{E}}(L_y|T_y, U_y) = \{\hat{\theta}_{U0} + \hat{\theta}_{U1}T_y + \hat{\theta}_{U2}U_y\}\exp(\hat{\sigma}_U^2/2)$$

are the estimated average soil lead concentrations in year $Y$ for a structure built in year $y$. The right sides of these relations were used to partition the estimated concentrations into fractional contributions of background, paint and gasoline. For this purpose, the proportionality constants $\exp(\hat{\sigma}^2/2)$ for equations (3.6) and (3.7) are not needed.

## 4. Results.

*Parameter estimates, $\theta$'s and $\sigma$'s.* Shown in Table 1 are the estimates of the $\theta$ parameters along with bootstrap [16] and large-sample likelihood-based standard errors. The bootstrap standard errors were constructed from 100 bootstrap samples of the residuals, adding each sample to the fitted values to obtain a new response from which bootstrap estimates were computed. The residual means were all of the order $5 \times 10^{-9}$, so no mean adjustment was used. The agreement between the two types of standard errors seems quite good.

The three black lines superimposed on each of the plots of Figure 1 are the fitted log concentrations plus and minus one standard error computed by year from the bootstrap replications used for Table 1. Although the fitted log concentrations are plotted against year, year itself was not used as a predictor in the models. The dotted lines are lowess smooths of the plotted data. The fitted log concentrations $\log(\hat{L}_y)$ shown in Figure 1 give a good representation of average concentration over time, supporting the possibility that the models are accurate.

*Diagnostics.* The fitted models were checked using various diagnostic procedures, including Bonferroni t-tests for outliers, marginal model plots [13], Cook's Distance [10] for influential observations and the score test for heteroscedasticity [11]. Some minor deviations from the models were detected, including three outliers, each in a different data set, that were just significant at the 0.05 level. None of these deviations was found to have a notable impact on the results and, thus, to preserve the integrity of the data and the conclusions, no remedial actions were taken. Additionally, we compared the results in Table 1 with those from fitting robust M-estimators using the R [41] function "nlrob" in package *robustbase*, obtaining very good agreement. One highly influential but not necessarily outlying sample was detected in the Minnesota foundation data. This sample is discussed later.

On balance, the models provide a good fit to the data, and indicate that the cumulative exposure variables are effectively taking the place of the year in which a structure was built.

It was reported for the National Survey that 95 percent of soil lead measurements would be within a factor of 2.7 of the true concentration [48]. This means that the measurement standard deviation is about 0.5 for the log-transformed lead concentrations. Comparing this with estimated standard deviations, $\sigma$ in Table 1 indicates that a substantial portion of the variation around the mean could be attributable to measurement error in lead concentration $L_y$. We found no information about measurement error for Minnesota.

TABLE 1
*Estimated rate parameters (Est.), standard errors (S.E.) and bootstrap standard errors (B.S.E.) from equation (3.6) for Minnesota and equation (3.7) for the US for both foundation and yard samples*

| | MN | | | US | | |
|---|---|---|---|---|---|---|
| Term | Est. | S.E. | B.S.E. | Est. | S.E. | B.S.E. |
| Foundations | | | | | | |
| $\theta_0$ (Background) | 15.03 | 9.76 | 9.85 | 9.71 | 6.22 | 6.21 |
| $\theta_{\cdot 1}$ (Paint) | 200.64 | 24.03 | 24.39 | 154.14 | 34.40 | 33.40 |
| $\theta_{\cdot 2}$ (Gasoline) | 9.93 | 6.34 | 5.70 | 7.73 | 3.31 | 3.25 |
| $\sigma$ | 1.02 | | | 1.33 | | |
| Yards | | | | | | |
| $\theta_0$ (Background) | 23.65 | 8.33 | 8.65 | 8.15 | 4.63 | 4.43 |
| $\theta_{\cdot 1}$ (Paint) | 34.73 | 5.68 | 6.05 | 46.64 | 13.63 | 15.35 |
| $\theta_{\cdot 2}$ (Gasoline) | 9.74 | 3.56 | 3.54 | 7.58 | 2.16 | 2.14 |
| $\sigma$ | 0.85 | | | 1.21 | | |

TABLE 2
*95 percent likelihood intervals for the four background concentrations*

| Term | Lower limit | Estimate | Upper limit |
|---|---|---|---|
| Foundations | | | |
| $\theta_{M0}$ | 3.26 | 15.03 | 53.48 |
| $\theta_{U0}$ | 0 | 9.71 | 24.12 |
| Yards | | | |
| $\theta_{M0}$ | 10.09 | 23.65 | 45.92 |
| $\theta_{U0}$ | 0.29 | 8.15 | 18.31 |

*Inference on background* $\theta_0$. Because models (3.4) and (3.5) are nonlinear, the standard paradigm of taking an estimate plus and minus twice its standard error for an approximate 95 percent confidence interval is not necessarily appropriate. However, using confidence curves [12], we concluded that, for the present study, this standard paradigm is reasonable for the $\theta$'s associated with paint and gasoline, but not for the background $\theta_0$'s. Instead, in Table 2 we present 95 percent profile likelihood intervals for the background concentrations. For these intervals, the uncertainty is not symmetric around the estimate. This may be apparent particularly for Minnesota foundation samples. The bootstrap distribution for $\theta_0$ is right skewed, which is in qualitative agreement with Table 2.

The estimated background concentrations from Table 2 are somewhat larger for Minnesota than for the National Survey. The US Geological Survey estimated the concentration of naturally occurring lead in soil to have a national mean of 16 ppm [43]. Thus, taking the uncertainty into account, all background estimates are consistent with the national average.

It has been estimated that about 30 percent of lead in burned gasoline spread beyond the surrounding region and contributed to continental and global polution [46]. This type of blanket contamination cannot be distinguished from the background with the available data and consequently will be reflected by an increase in $\theta_0$. Because the estimated background concentrations in Table 2 are consistent with the average background concentration for the US, blanket aerosol contamination was not found to be a significant source of elevated soil lead. This finding is consistent with Tiller et al. [46] who concluded that the general level of such contamination is "...much too low to affect human health."

In addition to the Minnesota data shown in Figure 1, the MPCA collected 41 samples that were designated as being from parks. Assuming that these samples were well removed from the nearest structure, the associated lead concentrations should reflect background and blanket contamination. The average lead concentration in the 41 park samples is 26.7 ppm with a standard error of 6.5. The average concentration for parks is thus well within sampling error when compared to the estimated background parameters for Minnesota in Table 2.

*Inference on the paint rate $\theta_{\cdot 1}$.* The cumulative exposure predictor $T_y$ for paint shown in Figure 3 changes relatively fast until about 1945 when it begins to level off. Consequently, samples from structures built before 1945 are relatively important for estimating the paint rates. Both the US and Minnesota studies contain a substantial number of samples from residences built before 1945 and as a result, the estimated rates $\theta_{\cdot 1}$ for paint are relatively well determined, all being larger than about 3.5 times their respective standard errors.

If lead from lead-based paint contributes substantially to the lead concentrations observed in foundation soil, then it is reasonable to suppose that paint rates for foundations would be noticeably larger than paint rates for yard samples. The results in Table 1 are consistent with this supposition.

*Inference on the gasoline rate $\theta_{\cdot 2}$.* Of the four estimated rate parameters for gasoline, all exceed twice their standard errors except for Minnesota foundations where $\hat{\theta}_{M2} = 9.93$ is only about 1.57 times its standard error. In addition, using Cook's Distance, it was found that $\hat{\theta}_{M2}$ is highly influenced by the newest structure in Figure 1(a), the one with the smallest value of $\log(L_y)$. With that sample removed, $\hat{\theta}_{M2} = 2.77$ with a standard error of 9.65. In effect, a substantial portion of the information on $\theta_{M2}$ rests with a single sample. This situation is caused in part by the Minnesota sampling plan. The cumulative concentration $M_y$ shown in Figure 3 is relatively constant until about 1945 when it begins to decrease rapidly. Consequently, samples from newer structures are relatively important for estimating the gasoline component. There are few residences built after the 1940's in the Minnesota foundation data, causing those present to be highly influential. Because there is no statistical or other evidence to indicate that the influential sample is anomalous, it was left in the data as observed. The other three gasoline rates are relatively well determined.

The estimated rates for gasoline are larger for Minnesota than the US. These differences, which are within sampling error, might be due in part to the inclusion of rural residences in the National Survey but not in the Minnesota study.

Under the aerosol hypothesis for accumulation of lead from leaded gasoline, we might expect the estimated gasoline rates for foundation samples to be noticeably larger than those for yard samples. However, in Table 1 the estimated gasoline rates for foundations are quite similar to those for yards, a finding that does not sustain the hypothesis.

Some extra caution should be used when comparing the estimates in Table 1 because the sampling protocols differ for the two studies and the Minnesota cumulative exposure predictor $M_y$ for gasoline differs from the US cumulative exposure predictor $U_y$. Measurement error in the predictors can also cause difficulties with interpretation because it can produce a bias in the results of Table 1, as discussed later in Section 5. In addition, it is not advisable to contrast paint and gasoline contributions by comparing the estimated rates in Table 1. But information on the relative importance of these sources can be gained by using estimated fractional contributions.

*Source apportionment.* Figures 4(a)–4(d) show the estimated fractional contributions (EFC) by year $y$, study and sample type of the three sources background, paint and gasoline. The two curves shadowing each EFC curve represent plus and minus one bootstrap standard error (BSE) computed by year from the bootstrap replications used for Table 1. For each point on an abscissa, the ordinates of the three EFC curves are positive and sum to one. In this way each of the three EFC curves give the fractional contribution to the estimated concentration $\hat{L}_y$ of the indicated source by year built. For instance, consider a Minnesota foundation sample from a structure built in 1952, the abscissa value at which the EFC curves for paint and gasoline cross in Figure 4(a). For such a sample, the contributions of paint and gasoline are about the same on the average, each being responsible for 42.4 percent of the estimated concentration $\hat{L}_{1952}$. The background concentration contributes 15.2 percent. From Table 1 the background concentration is estimated as 15.03 ppm and, thus, the total average concentration is about 99 ppm, consistent with the fitted line in Figure 1(a).

The BSE curves show that the paint contribution is relatively well determined in all four settings, but the standard errors for yards are somewhat larger than those for foundations, as might be expected. Since the ordinates of the three EFC curves sum to 1, the error for the EFC of gasoline plus background must be the same as that for paint. Thus, except for US yards, the results in Figure 4 indicate that the contributions of gasoline and background are difficult to distinguish for some years.

The EFC curves in Figure 4 can always be interpreted as the fractional contributions of background, paint and gasoline to the estimated mean concentration $\hat{E}(L_y|G_y, P_y)$ in year $y$. But to be interpreted unambiguously as source apportionment curves, equations (2.5) and (2.6) should hold to reasonable approximations when evaluated in terms $P_y \propto T_y$ and $G_y \propto U_y$. Consider the interpretation of equation (2.5); the interpretation of equation (2.6) is similar. The cumulative concentration $\rho_y$ due to paint will likely be statistically related to $T_y$ and to $U_y$ because both predictors decrease over time. However, if we know $T_y$, the national consumption of lead in lead-based paint in year $y$, then the national consumption of lead in leaded gasoline $U_y$ should furnish little or no *additional* information about the average level of $\rho_y$: $\mathrm{E}(\rho_y|T_y, U_y) = \mathrm{E}(\rho_y|T_y)$. If this interpretation is reasonable, then the EFC curves in Figure 4 can be interpreted reasonably as source apportionment curves. It is possible to imagine scenarios in which $\mathrm{E}(\rho_y|T_y, U_y) \neq \mathrm{E}(\rho_y|T_y)$, but none seem realistic to us in the present context.

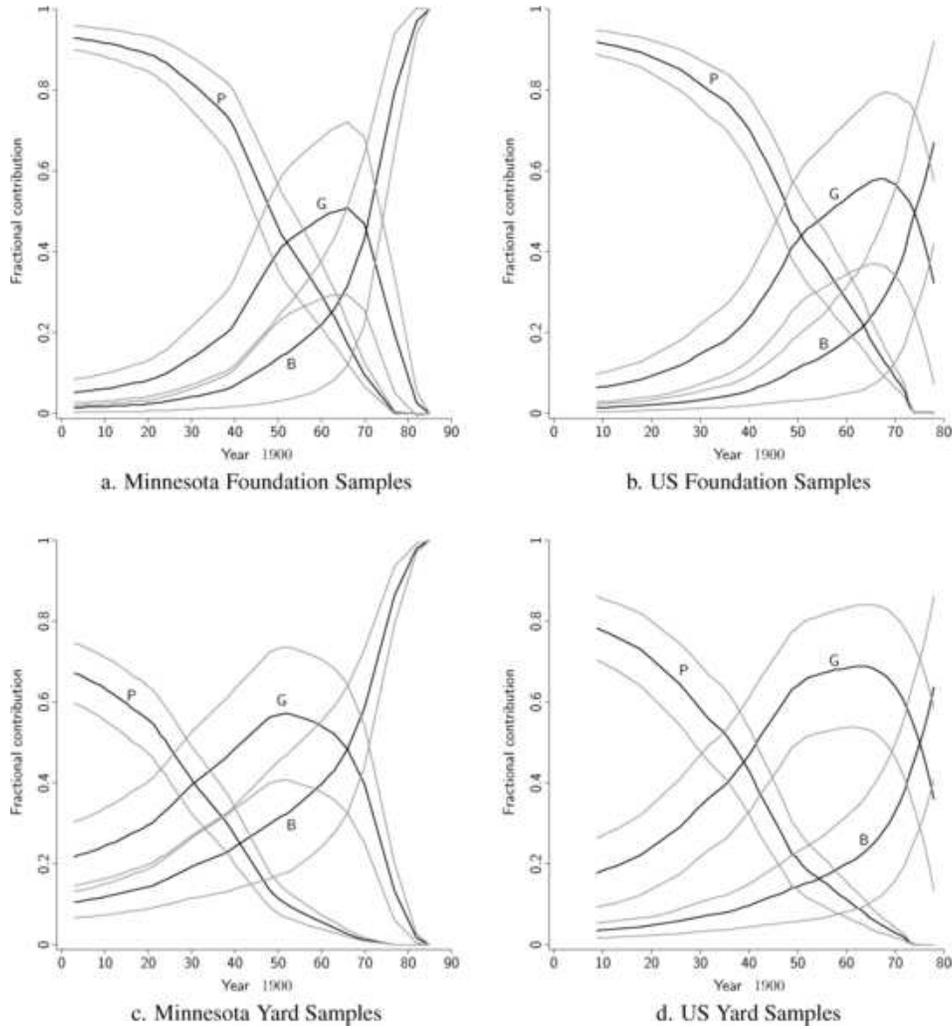

Fig. 4. *Fractional contributions of paint (P), gasoline (G) and background (B) to foundation and yard samples for Minnesota and US data. Estimated contributions are given by the black lines; gray lines are plus and minus one bootstrap standard error.*

The use of lead-based paint in the US was at its highest point in the 1920's and it declined markedly starting in the 1940's. Its use in residential paints was virtually banned in the late 1970's. Figures 4(a) and 4(b) both indicate that, on the average, lead from lead-based paint accounts for at least 80 percent of the lead in foundation soils from residences built before about 1925. For residences built after 1925, the fraction of lead from paint decreases substantially, dropping below the contribution of lead from gasoline for residences built after about 1950 and below the background level in the 1960's. Similarly, aerosol lead from gasoline is estimated to account for the greatest percentage of foundation lead for homes built between 1950 and 1970. The interpretation of Figures 4(c) and 4(d) is similar, except that the percentage of lead from leaded gasoline is estimated to be uniformly greater for yard samples than for foundation samples. Note also that the year-built interval in which lead from leaded gasoline accounts for the greatest percentage of soil lead is noticeably greater for yard samples than for foundation samples.

Fergusson and Schroeder [19] studied house dust in Christchurch, New Zealand. They concluded in part that for homes built between 1920 and 1940 and containing lead paint, about 45 percent of lead in house dust came from lead-based paint and about 50 percent came from leaded gasoline. These findings agree well with those for yard samples in Figure 4.

**5. Model extensions.** In this section we describe extensions of model (3.5) that facilitates discussion of measurement error and structure-specific fractional losses of lead. Measurement error is considered first. The motivation is primarily in terms of lead-based paint, but parallel reasoning applies to leaded gasoline.

Recall from equation (3.1) that we have used $T_y = \sum_y^Y w_i$ as a proportional replacement for $P_y$. If the $w$'s are subject to nonnegligible measurement error, then a more accurate version of equation (3.1) is

$$\mathrm{E}(\rho_y | T_y) = \theta_1 \sum_{i=y}^{Y} w_i \epsilon_i = \theta_1 T_y \Delta_y^{(p)},$$

where the $\epsilon_i \in (0, \infty)$ are independent random variables and

$$\Delta_y^{(p)} = \frac{\sum_{i=y}^{Y} w_i \epsilon_i}{\sum_{i=y}^{Y} w_i}$$

is a weighted average of the $\epsilon_i$'s. Terms of the form $w\epsilon$ allow for error in the measurement of white lead pigment $w$. We have used proportional errors since the $w_i$'s vary several orders of magnitude over the time period involved, making additive errors problematic. We then have

$$E(\Delta_y^{(p)}) = E(\epsilon),$$
$$\text{Var}(\Delta_y^{(p)}) = \{c_y^2(w) + 1\}\sigma_\epsilon^2/A_y,$$

where $A_y = Y - y + 1$ is the structure age, $c_y(w)$ is the sample coefficient of variation for $w_y, \ldots, w_Y$ and $\sigma_\epsilon^2 = \text{Var}(\epsilon)$. The covariance for structures built in years $y'$ and $y$, $y' < y$, is

$$\text{Cov}(\Delta_{y'}^{(p)}, \Delta_y^{(p)}) = \text{Var}(\Delta_y^{(p)}) \frac{\sum_{i=y}^{Y} w_i}{\sum_{i=y'}^{Y} w_i}.$$

The dependence of the $\Delta_y^{(p)}$'s is due to the sharing of measurement errors by structures of different ages.

Repeating the above argument for gasoline leads to an analogous independent error $\Delta_y^{(g)}$ and to the following expanded model for the National survey data:

$$(5.1) \qquad \log(L_y) = \log\{\theta_{U0} + \theta_{U1} T_y \Delta_y^{(p)} + \theta_{U2} U_y \Delta_y^{(g)}\} + \varepsilon.$$

We see from this result that sufficiently large biases, $E(\Delta^{(p)}) \neq 1$ and $E(\Delta^{(g)}) \neq 1$, in the measurements of $w$ and $u$ can in turn produce biased estimates of $\theta_{U1}$ and $\theta_{U2}$ and have an effect on all of the previous results, including the apportionment plots in Figure 4. We have no reason to suspect that any biases are present. Assuming no bias in the measurements of $w$ or $u$, $E(\Delta_y^{(p)}) = E(\Delta_y^{(g)}) = 1$, we can get a rough idea about the effects of measurement error by reasoning as follows. Under model (3.5), $\hat{\sigma}_U^2$ (see Table 1) is an estimator of the error variance $\text{Var}(\varepsilon)$. However, under model (5.1), $\hat{\sigma}_U^2$ will tend to be inflated by any bias in the mean function and by measurement error, and then $E(\hat{\sigma}_U^2) > \text{Var}(\varepsilon)$. An estimate $\widehat{V}_\varepsilon$ of $\text{Var}(\varepsilon)$ that is free of the effects of measurement error can be constructed by pooling the intra-year sample variances of $\log(L_y)$. The ratio $\hat{\sigma}_U^2/\widehat{V}_\varepsilon$ then provides an estimator of the excess variation due to measurement error and generally lack of fit. This ratio, along with its bootstrap standard error from the samples of Table 1, is $0.97 \pm 0.04$ for both US foundations and yards. For the Minnesota data this ratio is $1.17 \pm 0.10$ for foundations and $1.13 \pm 0.05$ for yards. There is some indication of excess variation for the Minnesota data, but on balance measurement error does not seen worrisome. A finer analysis could mitigate this first conclusion. We conjecture that the error $\varepsilon$, which depends on many different environmental components, dominates measurement error.

Our development in Section 2 led to parameterization in terms of the average yearly fractional loss of lead for a structure. An alternative modeling strategy could be based on representing loss fraction as a random effect. This route parallels that for measurement error, leading to the new representation $\mathrm{E}(\rho_{yj}|T_y) = \theta_1 T_y \Delta_{yj}^{(p)}$, where

$$\Delta_{yj}^{(p)} = \frac{\sum_{i=y}^{Y} w_i \epsilon_i \delta_{ij}}{\sum_{i=y}^{Y} w_i}.$$

All terms here are as defined previously, except $\delta_{ij} \in (0, \infty)$ is the loss of lead for the $j$th structure in year $i$ measured relative to the average loss. For example, $\delta_{ij} = 0.5$ means that in year $i$ structure $j$ lost only half of that for an average structure.

**6. Discussion.** Ideally we would have liked to apportion soil lead concentration $L_y$ locally in terms of equation (2.1) so that the fractions of contamination from background, paint and gasoline are simply $\beta/L_y$, $\rho_y/L_y$ and $\gamma_y/L_y$. Since $\rho_y$ and $\gamma_y$ are not estimable with current technology, we chose instead to base apportionment on the conditional average $\mathrm{E}(L_y|G_y, P_y) = \mathrm{E}(\beta) + \mathrm{E}(\rho_y|P_y) + \mathrm{E}(\gamma_y|G_y)$, which depends on the reasonableness of equations (2.5) and (2.6).

Our results, including the apportionments in Figure 4, apply to averages over residential populations as reflected by the Minnesota and US sampling plans. We view the rather striking agreement between the results of the Minnesota and US analyses as a strong point in favor of our approach. Our results agree at least qualitatively with many findings from past studies. The aerosol hypothesis is a notable exception to this agreement. For instance, we estimate from Figure 4 that, on the average, lead from lead-based paint is the primary contributor to foundation contamination for structures built before 1940, while lead from leaded gasoline is the dominant contributor for structures built between 1950 and 1970. After that, contamination from paint and lead are both estimated to be below background. Our findings prompted the following modified aerosol hypothesis: Relatively heavy particles in lead aerosols are deposited on or near the road bed, while relatively light particles are largely carried around structures by air currents, resulting in blanket contamination and general atmospheric pollution. Only particles of intermediate mass are carried by air currents and have an affinity for structures.

The ideas underlying our apportionment method are similar to the pioneering work of Mosteller and Lagakos [20, 30, 35] on assigned shares for compensation in radiation-related cancers. See [39] for a discussion of the impact of this work. Briefly, letting $r(t|d)$ denote the cancer incidence rate for an age $t$ population exposed to a dose $d$ of radiation, the assigned share of cancers due to exposure is $\mathrm{AS} = \{r(t|d) - r(t|0)\}/r(t|d)$. Under a model of constant excess risk, $r(t|d) = r(t|0) + f(d)$ for some dose-response function $f$, we have $\mathrm{AS} = f(d)/r(t|d)$ which is similar in spirit to the fractional contributions used here. Our doses are represented by $G_y$ and $P_y$, time is structure age and $\mathrm{E}(L_y|G_y, P_y)$ plays the role of the risk. Mosteller and Lagakos were concerned with providing a scientific basis for compensating cancer victims who were exposed to radiation. Similar but distinct issues may arise in the context of lead poisoning where, in the absence of evidence to the contrary, culpability may be seen to rest with the obvious suspects. Given total compensation, the issue may center on who should provide it and in what proportions. Similar questions may arise in lead abatement litigation.

Because soil lead concentration can vary substantially from residence to residence as depicted in Figure 1, an individual residence or community might not conform closely to the average behavior described here. Model (2.7) can be modified straightforwardly to include local, issue-specific variables. In a community located near a point-source emitter, information on distance to source and cumulative emissions or production can be added as terms inside the log term on the right side of (2.7). Paint condition may be an issue in some litigation. The Minnesota data include qualitative information on paint condition as excellent, good, average, fair or poor. This information can be incorporated into the model by adding indicator variables.

**Acknowledgments.** The analysis of the Minnesota data was supported in part by the Ethyl Corporation. The analysis of the US data was supported by a grant from the National Science Foundation and an in-house grant from the University of Central Florida. The authors are grateful to Jerry Roper for helpful discussions throughout the investigation and for providing yearly data for the construction of the cumulative exposure predictors. We are grateful also to Douglas Hawkins for feedback on model development and to Stephen Fienberg for comments on a preliminary version of this article. The Associate Editor and Referee provided many suggestions that resulted in significant improvements. The data base for the Minnesota study was managed by the Minneapolis urban planning firm Dahlgren, Shardlow and Uban, Inc., including the transfer of paper records to a computer format and the search of the public record for the year built.

## REFERENCES

[1] ASCHENGRAU, A., BEISER, A., BELLINGER, D., COPENHAFER, D. and WEITZMAN, M. (1994). The impact of soil lead abatement on urban children's blood lead levels: Phase II results from the Boston lead-in-soil demonstration project. *Environ. Res.* **67** 125–148.

[2] BROWN, K. W., MULLINS, J. W., RICHITT, E. P. JR., FLATMAN, G. T. and BLACK, S. C. (1985). Assessing soil-lead contamination in Dallas, Texas. *Environ. Monitoring Assess.* **5** 137–154.

[3] BROWNE, F. L. and LAUGHNAN, D. F. (1953). Effect of coating thickness on the performance of house paints under different programs of maintenance. *Official Digest* **March** 137–159.

[4] BURGOON, D. A., BROWN, S. F. and MENTON, R. G. (1995). Literature review of sources of elevated soil-lead concentrations. *Lead in Paint, Soil and Dust: Health Risks, Exposure Studies, Control Measures, Measurement Methods, and Quality Assurance* (M. E. Beard and D. D. A. Iske, eds.) 76–91. American Society for Testing and Minerals, Philadelphia.

[5] CARROLL, R. J. and RUPPERT, D. (1988). *Transformation and Weighting in Regression.* Chapman and Hall, London. MR1014890

[6] CHANEY, R. L. and MIELKE, H. W. (1986). Standards for soil lead limitations in the United States. *Trace Substances in Environ. Health* **20** 357–377.

[7] CHILLRUD, S. N., BOPP, R. F., SIMPSON, H. J., ROSS, J. M., SCHUSTER, E. J., CHAKY, D. A., WALSH, D. N., CHOY, C. C., TROLLEY, L. and YARME, A. (1999). Twentieth century atmospheric metal fluxes into central park lake, New York City. *Environ. Sci. Tech.* **33** 657–652.

[8] CHRISTENSEN, W. F. and GUNST, R. F. (2004). Measurement error models in chemical mass balance analysis of air quality data. *Atmosph. Environ.* **38** 733–744.

[9] CLARK, S., BORNSCHEIN, R., SUCCOP, P. and PEACE, B. (1988). The Cincinnati soil-lead abatement demonstration project. *Lead in Soil: Issues and Guidelines* (B. E. Davies and B. G. Wixson, eds.) 287–300. Science Reviews Ltd., Northwood.

[10] COOK, R. D. (1977). Detection of influential observation in linear regression. *Technometrics* **19** 15–18. MR0436478

[11] COOK, R. D. and WEISBERG, S. (1983). Diagnostics for heteroscedasticity in regression. *Biometrika* **70** 1–10. MR0742970

[12] COOK, R. D. and WEISBERG, S. (1990). Confidence curves in nonlinear regression. *J. Amer. Statist. Assoc.* **85** 544–551.

[13] COOK, R. D. and WEISBERG, S. (1997). Graphics for assessing the adequacy of regression models. *J. Amer. Statist. Assoc.* **92** 490–499. MR1467843

[14] DAVIES, B. E. (1988). Lead in soil: Its sources and typical concentration. In *Lead in Soil: Issues and Guidelines* (B. E. Davies and B. G. Wixson, eds.) 65–72. Science Reviews Ltd., Northwood.

[15] DAVIS, A. P. and BURNS, M. (1999). Evaluation of lead concentration in runoff from painted structures. *Water Reservation* **33** 2949–2958.

[16] EFRON, B. and TIBSHIRANI, R. J. (1993). *An Introduction to the Bootstrap.* Chapman and Hall, London. MR1270903

[17] ETHYL CORPORATION (1984). *Yearly Report of Gasoline Sales by State.* 2 Houston Center, Suite 900, Houston, Texas, 77010.

[18] FARRELL, K. P., BROPHY, M. C., CHISOLM, J. J. JR., RHODE, C. A. and STRAUSS, W. J. (1998). Soil lead abatement and children's blood lead levels in an urban setting. *Amer. J. Public Health* **88** 1837–1893.

[19] FERGUSSON, J. E. and SCHROEDER, R. J. (1985). Lead in house dust of Christchurch, New Zealand: Sampling, levels and sources. *Sci. Total Environ.* **46** 61–72.

[20] FIENBERG, S. E. and HOAGLIN, D. C. (2006). *Selected Papers of Frederick Mosteller.* Springer, New York.

[21] FRANCEK, M. A. (1992). Soil-lead levels in a small town environment: A case study from Mt. Pleasant, Michigan. *Environ. Pollution* **76** 251–257.

[22] GRANEY, J. R., DVONCH, J. T. and KEELER, G. J. (2004). Use of multi-element tracers to source apportion mercury in south Florida aerosols. *Atmosph. Environ.* **38** 1715–1726.

[23] HARRISON, R. M. and LAXEN, D. P. H. (1981). *Lead Polution Causes and Control*. Chapman and Hall, London.
[24] HENRY, R. C., LEWIS, C. W., HOPKE, P. K. and WILLIAMSON, H. J. (1984). Review of receptor model fundamentals. *Environ. Res.* **67** 125–148.
[25] HOPKE, P. K. (1985). *Receptor Modeling in Environment Chemistry*. Wiley, New York.
[26] HUNTZICKER, J. J., FRIEDLANDER, S. K. and DAVIDSON, C. (1975). Material balance for automobile-emitted lead in Los Angeles basin. *Environ. Sci. and Technol.* **9** 448–457.
[27] INTERNATIONAL LEAD ZINC RESEARCH ASSOCIATION, INC. (1998). *Lead in Gasoline*: *Environmental Issues*. P.O. Box 12036, Research Triangle Park, North Carolina, 27709-2036.
[28] KAMINSKI, M. D. and LANDSBERGER, S. J. (2000). Heavy metals in urban soils of east St. Louis, IL, Part I: total concentration of heavy metals in soils. *J. Air Waste Manag. Assoc.* **50** 1667–1679.
[29] KIRT-OTHMER (1992). *Encyclopedia of Chemical Technology*, 4th ed. Wiley, New York. 342–350.
[30] LAGAKOS, S. W. and MOSTELLER, F. (1986). Assigned shares in compensating for radiation-related cancers. *Risk Analysis* **6** 345–357.
[31] MASKALL, J., WHITEHEAD, K. and THORNTON, I. (1995). Heavy metal migration in soils and rocks at historical smelting sites. *Environ. Geochem. Health* **17** 127–138.
[32] MIELKE, H. W. (1991). Lead in residential soils: Background and preliminary results of New Orleans. *Water, Air, Soil Pollution* **57–58** 111–119.
[33] MIELKE, H. W. (1997). Leaded dust in urban soil shown to be greater source of childhood lead poisoning than leaded paint. *Lead Perspect.* **March/April** 28–31.
[34] MIELKE, H. W., ANDERSON, J. C., BERRY, K. J., MIELKE, P. W., CHANEY, R. L. and LEECH, M. (1983). Lead concentrations in the inner-city soils as a factor in the child lead problem. *Am. J. Public Health* **73** 1366–1369.
[35] MOSTELLER, F. (1987). Compensating for radiation-related cancers by probability of causation or assigned shares. *Bulletin of the ISI* **52** 571–577.
[36] MINNESOTA POLUTION CONTROL AGENCY (1987). *Soil Lead Report to the Minnesota State Legislature*.
[37] NRIAGU, J. O. (1990). The rise and fall of leaded gasoline. *Sci. Total Environ.* **92** 13–28.
[38] PAGE, A. L. and GANJE, T. J. (1970). Accumulations of lead in soils for regions of high and low motor vehicle traffic density. *Environ. Sci. and Technol.* **4** 140–142.
[39] PARASCANDOLA, M. (2002). The NIH radioepidemiologic tables and compensation fr radiation-induced cancer. *Isis* **93** 559–584.
[40] PARK, E. S., GUTTORP, P. and HENRY, R. C. (2001). Multivariate receptor modeling for temporally correlated data by using MCMC. *J. Amer. Statist. Assoc.* **96** 1171–1183. MR1946572
[41] R DEVELOPMENT CORE TEAM (2006). *R*: *A Language and Environment for Statistical Computing*. R Foundation for Statistical Computing, Vienna, Austria. ISBN 3-9000510-07-0. Available at http://www.R-project.org.
[42] SCHMITT, M. D. C., TRIPPLER, D. J., WACHTLER, J. N. and LUND, G. V. (1988). Soil-lead concentrations in residential Minnesota as measured by ICP-AES. *Water Air Soil Pollution* **39** 157–168.
[43] SHACKLETTE, H. T. and BOERNGEN, J. G. (1984). Element concentrations in soils and other surficial materials of the conterminous United States. U.S. Geological Survey Professional Paper 1270.

[44] SINGER, M. J. and HANSON, L. (1969). Lead accumulation in soils near highways in the Twin Cities metropolitan area. *Soil Science Society of America Proceedings* **33** 152–153.
[45] TER HAAR, G. and ARONOW, R. (1974). New information on lead in dirt and dust as related to the childhood lead problem. *Environ. Health Perspect.* **7** 83–89.
[46] TILLER, K. G., SMITH, L. H., MERRY, R. H. and CLAYTON, P. M. (1987). The dispersal of automotive lead from metropolitan Adelaide into adjacent rural areas. *Aust. J. Soil Res.* **25** 155–166.
[47] U.S. DEPARTMENT OF HOUSING AND URBAN DEVELOPMENT (2001). *National Survey of Lead and Allergens in Housing*, Final Report, **1**: *Analysis of Lead Hazards*.
[48] U.S. ENVIRONMENTAL PROTECTION AGENCY (1993). *Data Analysis of Lead in Soil and Dust*. EPA 747-R-93-011.
[49] U.S. ENVIRONMENTAL PROTECTION AGENCY (1996). *Urban Soil Lead Abatement Demonstration Project* **I**. EPA 600-P-93-001aF.
[50] U.S. ENVIRONMENTAL PROTECTION AGENCY (1998). *Sources of Lead in Soil*: *A Literature Review*. EPA 747-R-98-001a.
[51] U.S. ENVIRONMENTAL PROTECTION AGENCY (2001). *Lead*; *Identification of Dangerous levels of Lead*; *Final Rule* **40** *CFR Part* **745** 1205–1240.
[52] ZIMDAHL, R. L. and SKOGERBOE, R. K. (1977). Behavior of lead in soil. *Environ. Sci. and Technol.* **11** 1202–1207.

SCHOOL OF STATISTICS
UNIVERSITY OF MINNESOTA
MINNEAPOLIS, MINNESOTA 55455
USA
E-MAIL: dennis@stat.umn.edu

DEPARTMENT OF STATISTICS AND ACTUARIAL SCIENCE
UNIVERSITY OF CENTRAL FLORIDA
ORLANDO, FLORIDA 32816-2370
USA
E-MAIL: lni@mail.ucf.edu